\documentclass{dsc}

\usepackage[utf8]{inputenc} 
\usepackage{amsmath}
\usepackage{amsfonts}
\usepackage{amssymb}
\usepackage{lmodern}
\usepackage{multirow}
\usepackage{xcolor}
\usepackage{float} % Required for the H float option
\usepackage{wrapfig} % Required for the wrapfigure environment

\ifpdf \usepackage[pdftex]{graphicx} \pdfcompresslevel=9
\else \usepackage[dvips]{graphicx} \fi

\usepackage{acronym}
\usepackage{svg}
\title[Requirement Identification for Traffic Simulations]{Requirement Identification for Traffic Simulations in Driving Simulators}

\author[Tarlowski]
	{\textbf{Sven Tarlowski and Lutz Eckstein}}
\begin{document}

\setcounter{page}{1}

\maketitle

\begin{affiliation}
Institute for Automotive Engineering (ika) RWTH Aachen University, sven.tarlowski@ika.rwth-aachen.de
\end{affiliation}

\begin{abstract}
% (DSC-Abstract)The abstract is written in Arial, size 10. The paragraph is justified, one-column (full page width). The abstract is limited to 200 words. It is followed by up to 5 keywords.
This paper addresses the challenge of ensuring realistic traffic conditions by proposing a methodology that systematically identifies traffic simulation requirements.
Using a structured approach based on sub-goals in each study phase, specific technical needs are derived for microscopic levels, agent models, and visual representation.
The methodology aims to maintain a high degree of fidelity, enhancing both the validity of experimental outcomes and participant engagement.
By providing a clear link between study objectives and traffic simulation design, this approach supports robust automotive development and testing.
\end{abstract}


% \section*{Introduction (DSC-Title 1)} \subsection*{Subtitle (DSC-Title 2)} \subsubsection*{Paragraph (DSC-Title 3)}
\section*{Introduction}

Driving simulators are essential tools in the automotive industry, facilitating simulation-based development and testing. 
They provide a controlled, reproducible environment for evaluating systems in the early stages of development, thereby enabling valuable feedback collection \citep{Winner.2024}.

To ensure that the feedback collected within driving simulator studies is valid, a high fidelity of the driving simulator is required.
There are multiple aspects that influence the fidelity of a driving simulator.
One of the aspects is the simulation of the surrounding traffic. \citep{Fischer.2015}

In order to determine the fidelity requirements for the traffic simulation needs to be defined.
This is a challenging task as there are various designs of traffic simulations and each study has its own specific requirements.
In order to derive requirements for the traffic simulation a systematic approach is needed.

This paper aims to propose a methodology for identifying requirements for traffic simulation in driving simulator studies based on the study design.

\section*{State of the Art}
\label{sec:state_of_the_art}
Designing a driving simulator study begins with defining the research question, as it determines the specific requirements and constraints of the simulator.
In turn, these requirements guide the selection of a suitable driving simulator and the structure of the study design.
Although each study design may differ according to its particular objectives, certain elements such as the familiarization, experimental trial and distraction phase are consistently found across most studies. \citep{Winner.2024}

High fidelity of the driving simulator is critical to ensure valuable study outcomes.
Multiple approaches have been proposed to validate driving simulators and their study designs, illustrating both essential components and methods for their evaluation \citep{Hock.2018}.
One significant aspects influencing overall fidelity is traffic simulation, which focuses on modeling the surrounding traffic participants \citep{Fischer.2015}.
Various frameworks exist in the literature, such as the co-simulation approach using the SUMO platform for realistic traffic \citep{Salles.2022}, or multi-level simulations (microscopic, mesoscopic, macroscopic) integrated with driver models \citep{Olstam.2008}.
Although these methods comprehensively address traffic behavior, they leave room to further explore how the traffic simulation can be tailored or adapted to specific study designs.
Establishing a stronger link between a study design requirements and the parameters of the traffic simulation could enhance the overall fidelity and relevance of driving simulator studies.

\section*{Methodology}

Despite the importance of traffic simulation in driving simulators, there remains a lack of a systematical approach to deriving requirements from an existing study design.
The following methodology addresses this gap by mapping each phase of the study design to specific traffic simulation requirements.

As described in the state of the art, each study has a research question that needs to be answered.
This research question is defined  by the study goal. 
A study design is defined which should help to answer the research question and fulfill the study goal.
The study design will include different phases, which each serve a different purpose.
To ensure that the overall study goal is achieved, each phase will contribute by setting its own specific goals, so called sub-goals.
Each phase can have multiple sub-goals.

Each sub-goal will influence the technical design of the simulator and simulation environment.
Based on the sub-goals, technical requirements for the traffic simulation can be derived.
To simplify this derivation process, the traffic simulation can be subdivided into three categories:
(1) microscopic level, covering elements such as traffic density, speed, and interaction rates; 
(2) agent model, which governs individual vehicle behaviors like time-headway or lane-changing; and (3) visual aspects, defining how traffic participants appear (vehicle types, colors, etc.). 

\begin{figure}[h]
	\centering
	\includegraphics[width=\linewidth]{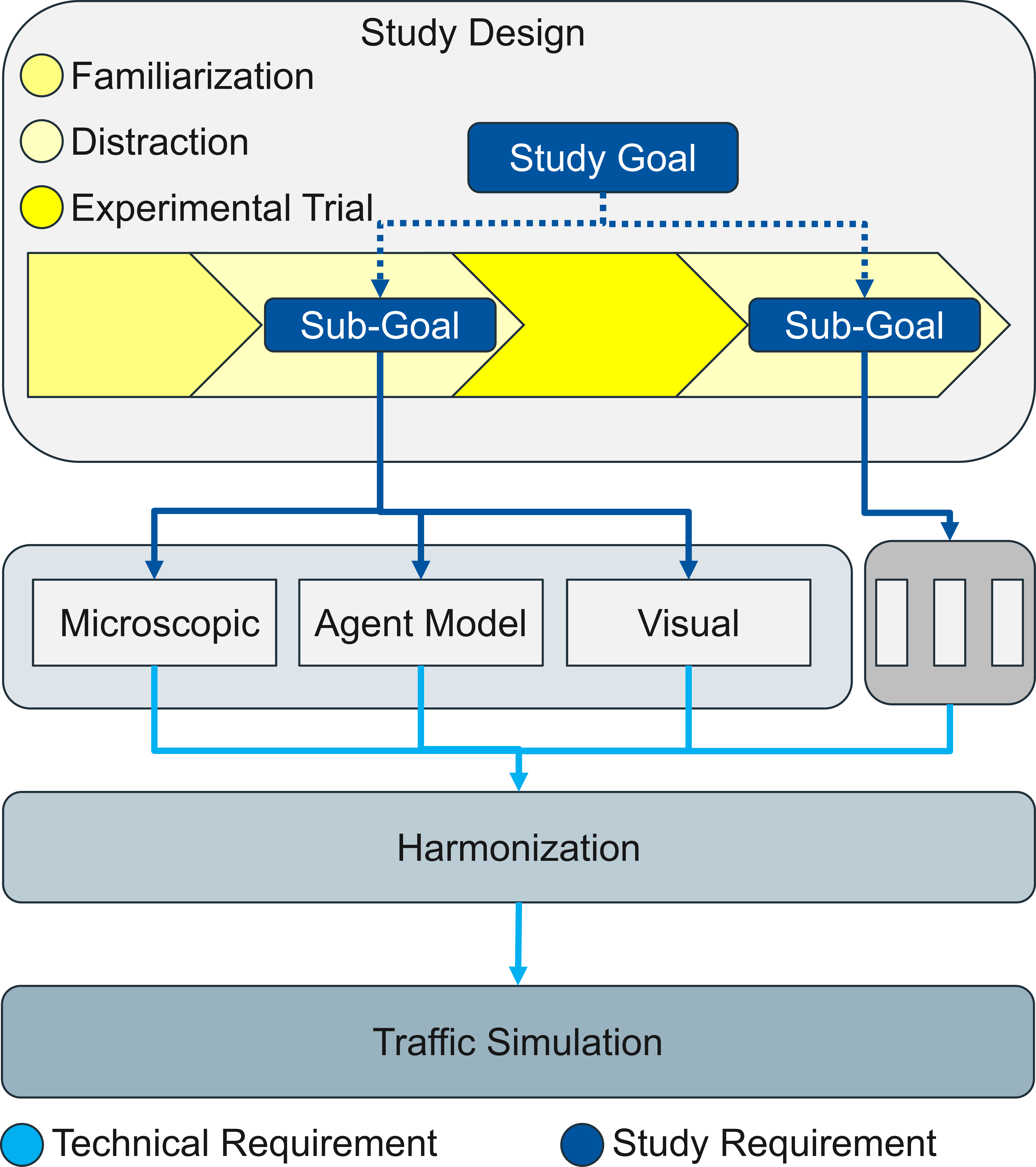}
	\caption{Connection between the study goals and the traffic simulation requirements} 
	\label{fig:design}
\end{figure}

Figure \ref{fig:design} shows the connection between the study goal and the technical requirements with the overall approach.
The study contains the study goal and the study design.
As an example, a study goal can be defined as \textit{Study Goal: Identifying  the driver response to a cut-in manoeuvre}.

For this study goal a study design with a familiarization phase, an experimental trial and a distraction phase is defined.
The distraction phase should separate  the familiarization from the tested scenario by letting the driver drive in normal traffic conditions.
In this example, the sub-goal within the distraction phase can be defined as \textit{Sub-Goal: The driver should drive in normal traffic conditions unaware of a potential manoeuvre}.
This sub-goal is then setting requirements for the traffic simulation.
The other phases can also set sub-goals for their specific purpose.

The requirement for the traffic simulation based on the sub-goals must be translated into technical requirements.
For the microscopic level the given sub-goal can be translated into \textit{the traffic density should be set to a predefined value which is constant for the whole study}.
This ensures that the driver is not aware of the cut-in manoeuvre by a change in traffic density.
For the agent model, it should be ensured that no critical events occure, which would lead to a change in the driver behavior.
Thus appropriate agent model parameters must be selected.
The visual aspects should be set to a realistic representation of the traffic.
Here, no type of vehicle should stand out from the others.

As there are multiple sub-goals this also leads to multiple requirements for the traffic simulation.
In a next step, the requirements for each phase need to be harmonized.
This can potentially lead to conflicts between the requirements.
In those case, either the derived sub-goal or the traffic simulation requirement needs to be adapted.
If this brings no solution, the study design needs to be adapted.

\section*{Results}

The presented methodology provides a structured approach to identify the requirements for traffic simulation in driving simulator studies.
In a first step, the study design and study goal is defined.
With each phase in the study design specific sub-goals are defined.
Those sub-goals are then used to derive the technical requirements for the traffic simulation.
If needed a harmonization of the requirements can be done.
This approach allows a systematical derivation of requirements.
Those requirements can then be used to setup the traffic simulation and to validate it.
Further investigations are needed to determine how these requirements can be harmonized and achieved through traffic simulation.

% Displays the references
\printbibliography

\begingroup
\small
\section*{Acknowledgement}
Funded by the European Union. Views and opinions expressed are however those of the author(s) only and do not necessarily reflect those of the European Union or European Climate, Infrastructure and Environment Executive Agency (CINEA). Neither the European Union nor the granting authority can be held responsible for them. 
\begin{figure}[h]
	\centering
	\includegraphics[width=0.79\linewidth]{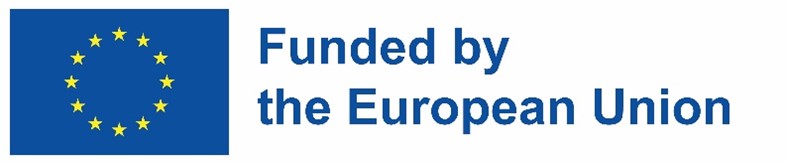}
\end{figure}
\endgroup
% \item The command \verb+\cite+ does not use put brackets around references: \cite{Dibs2008,Bailey2010}.
% \item The command \verb+\citep+ is used to put the references between brackets: \citep{Dibs2008,Bailey2010}.
% \item The command \verb+\citet+ puts brackets only around the publication year: \citet{Dibs2008,Bailey2010}.
\end{document}